\documentstyle[12pt]{article}
\def\kms{km s$^{-1}$}
\def\Msun{M_\odot}
\def\ha{H$\alpha$}

\begin{document}

\title{CENTRAL ROTATION CURVES OF SPIRAL GALAXIES}
\author{Y. SOFUE$^1$, Y. TUTUI$^1$, M. HONMA$^{1,3}$,\\
A. TOMITA$^2$, T. TAKAMIYA$^1$, J. KODA$^1$ \& Y. TAKEDA$^1$\\
{\it 1.Institute of Astronomy, Univ. of Tokyo, Mitaka, Tokyo 181-8588, Japan}\\
{\it 2. Faculty of Education, Wakayama University, Wakayama 640-8510, Japan}\\
{\it 3.National Astronomy Observatory,  Mitaka, Tokyo 181-8588, Japan}
}
\date{}
\maketitle

\begin{abstract}

We present high-resolution central-to-outer rotation curves for
Sb, SBb, Sc, and SBc galaxies.
We discuss their general characteristics, particularly their
central behavior, as well as dependencies on morphological
types, activity, and peculiarity.
The rotation curves generally show a steep nuclear rise, and high-velocity
central rotation, followed by a broad maximum in the disk and then by
a flat rotation due to the massive halo.
Since the central high velocity and steep rise are common to all 
massive galaxies, it cannot be due to  non-circular motions.
Disk rotation curves of barred galaxies show larger dispersion 
than those of normal galaxies which is probably due to non-circular motions.
Interacting galaxies show often perturbed outer rotation curves, 
while their central rotation shows no particular peculiarity.
Also, central activities, such as starbursts and AGN, appear to show no 
particular correlation with the property of rotation curves.
This would suggest that the central activities are
triggered by a more local effect than the global dynamical property.


\end{abstract}

\section{INTRODUCTION}

Rotation curves  are the principal tool to derive the axisymmetric
distribution of mass in spiral galaxies in the first-order approximation.
Rotation curves of galaxies in the disk and outer regions have been obtained
based on optical and HI-line spectroscopy (Rubin et al 1980, 1982;
Bosma 1981; Clemens 1985; Mathewson et al 1996; Persic and Salucci 1995;
Persic et al 1996; Honma and Sofue 1997).
These rotation curves have been used to estimate the mass distribution
in the disk and halo, particularly for the dark halo, using a
model-potential fits method (e.g., Kent 1987, 1992).
Persic et al (1996) have extensively studied the universal characteristics
of rotation curves in the outermost regions.
However, the inner rotation curves have been not thoroughly investigated 
yet in sufficient accuracy, not only because the concern in these studies 
has been on the distribution of mass in the outermost regions,
but also because of the difficulty in deriving the central rotation velocities.
The difficulty in measuring central rotation curves is mainly due to the 
lack of HI gas in the central regions, as well as due to contamination 
of bright bulge light, when photographic plates had been used.

In our recent papers, we have stressed that, in order to derive central
rotation curves, CO molecular lines are most convenient because of the high
concentration of molecular gas in the centers of many galaxies, 
the high angular and velocity resolutions in CO observations, and
the  negligible extinction even toward the nuclear dusty disk.
We have thus used high-resolution CO-line data to obtain
well-sampled inner rotation  curves for nearby galaxies
(Sofue 1996, 1997, Sofue et al 1997, 1998: Papers I to IV).
Recent CCD \ha\ line spectroscopy has also made 
available to us accurate rotation curves for the inner
regions, because of the large-dynamic range,
and the more precise subtraction of bulge continuum light for digital images,
in so far as the extinction is not very large:
The obtained optical position-velocity (PV) diagrams often show 
high-velocity central components
(Rubin et al 1997; Sofue et al 1998; Bertola et al 1998).
However, it is not yet known whether the central steep rise of 
rotation velocity is universal,  how it is correlated to other 
properties of galaxies, such as the Hubble types and central activity.

In this paper, we present high-accuracy rotation curves for Sb, SBb,
Sc and SBc galaxies, and discuss their general characteristics,
particularly for central rotation curves.
We also discuss their dependency on morphological
types, activity, and peculiarity.
Individual rotation curves are shown in the Appendix, and machine-readable
data are available by contacting the first author.
The rotation curve data, from the central to outer regions in high accuracy,
will be used in deriving the detailed mass and mass-to-luminosity
ratios in a separate paper (Takamiya and Sofue 1999a).

\section{CENTRAL-TO-OUTER ROTATION CURVES}

\subsection{Milky Way}

The rotation curve of the Milky Way Galaxy is shown in Figure 1a,
as reproduced from the literature (Clemens 1985; Honma and Sofue 1997).
Recently, it has been shown that the velocity dispersion
of stars within the central 10 pc increases toward the center, 
indicating the existence of a massive black hole
(Genzel et al 1997, 1998; Ghez et al 1998).
Assuming the existence of a black hole of $2.6\times 10^7\Msun$,
we have calculated a corresponding rotation curve, and combined it
with the existing CO data to obtain a rotation
curve of the Milky Way as shown in Figure 1b in logarithmic plot. 
Besides the Galaxy, evidence for nuclear massive black holes has been
accumulating for many other galaxies (Miyoshi et al 1995;
Richstone, et al. 1998), suggesting similar central rotation curves
for these galaxies (See section 3.3).

Thus, the rotation curve of our Galaxy can be described as having

{\parskip 0pt \parindent 0pt

(a) a high-density core, including the massive black hole, which causes a
non-zero velocity very close to the center.

(b) a steep rise within the central 100 pc.

(c) a maximum at radius about 300 pc, followed by a decline to a 
minimum at 2 kpc.

(d) a gradual rise from 2 kpc to the disk maximum at 6 kpc.

(e) a nearly flat outer region with a dip at 8 kpc, followed by a  second 
maximum at 15 kpc,  followed further by an outermost Keplerian decline.
}

\centerline {---- Fig. 1  -----}

\subsection{Well-Sampled Extended Rotation Curves}

Except for the Milky Way, it has been widely believed that the central 
rotation curves of most galaxies behave in a rigid-body fashion.
These rotation curves appear to behave differently from that of the
Milky Way's rotation in the inner disk and bulge regions.
The question may arise then as to whether the Milky Way really is an 
exception, or whether a similar rotation property is not showing
up in the currently published rotation curves of galaxies.

We have performed high-resolution CO-line observations of
galaxies to obtain PV diagrams, and combined them with
existing HI and optical rotation curves.
We have also obtained CCD spectroscopy in the \ha\
and [NII] line emissions of the central regions of galaxies.
We have applied the envelope-tracing method to derive rotation
curves from PV diagrams.
In Figure 2a we show the thus obtained most-completely sampled rotation
curves for all sample galaxies (Papers I - IV),
and in Figure 2b the same but for the central 5 kpc.
Figure 2c shows some rotation curves plotted against radii
normalized by the scale length radius of the exponential disk.

In deriving rotation curves, we have applied the envelope-tracing
method, which traces the terminal velocities in position-velocity
diagrams along the major axes, which is described in detail in Papers I and II.
This method has an advantage to estimate the rotation velocity in
the central regions more accurately than those used in the current
analyses (e.g., Warner et al 1972; Rubin et al 1980; Persic et al 1996), 
which gave intensity-weighted velocities. (See also section 4.2.)

In Figures 3a, b and c,  we show rotation curves for Sb, Sc,
and   barred (SBb and SBc) galaxies, respectively.
Figure 4 shows rotation curves for peculiar galaxies and
interacting galaxies.
Individual rotation curves are published in the literature as above
with detailed discussion, and displayed in Figure A1 of Appendix.
Parameters for individual galaxies are given in Table 1.

\centerline {--- Fig. 2   ---}

\centerline{ --- Fig. 3  ---}

\subsection{Sb Galaxies}

All Sb galaxies in   Figure 3a have rotation curves with
a very steep rise in the central 100-200 pc region, often associated
with a peak at radii $r \sim 100-300$ pc.
The rotation velocity, then, declines to a minimum at $r\sim 1 $ kpc,
and is followed by a gradual rise to a broad maximum at
$r \sim 2-7$ kpc, corresponding to the disk.
The outermost parts are usually flat, indicating the massive dark halo.
Some galaxies like the Milky Way show a declining outer
rotation (Honma and Sofue 1997a, b).
Thus, the rotation curves for Sb galaxies are essentially the 
same as that of the Milky Way Galaxy.

\subsection{Sc Galaxies}

Sc galaxies tend to have slower velocities than Sb, and the rotation
velocities are more spread from $\sim 100$ to $\sim 200$ \kms\
among the sample galaxies.
Massive Sc galaxies show a steep nuclear rise similar to Sb's,
while less-massive galaxies have a more gentle rise.
They also have a flat rotation until their outer edges.

\subsection{Barred SBb and SBc Galaxies}

In Figure 3 we compare  barred galaxies with non-barred galaxies.
There appears to be no particular difference in their general properties:
The rotation properties of the barred galaxies are almost the
same as those for non-barred galaxies of Sc and Sb types.
However, barred galaxies tend to show a larger-amplitude velocity 
variation with radius of about $\pm \sim 30 - 40$ \kms\ within 
the main disk at $R\sim 2 - 5$ kpc.
The large velocity variation may be due to barred potential of 
length of several kpc.
In order to clarify this and investigate the non-circular motions,
two-dimensional velocity fields are necessary, which is, however, 
out of the scope of this paper, and remains as a future subject.
On the other hand, normal galaxies usually show velocity variation
of about $\pm 10 - 20$ \kms\, mainly caused by spiral arms, 
except for a few cases.

\subsection{Peculiar and Interacting Galaxies}

Rotation curves for irregular galaxies NGC 660, NGC 3034, NGC 4631 
and NGC 4945 are shown in Figure 4a.
NGC 660 is a polar-ring galaxy with a warped galactic disk, while
the rotation curve from the disk to outer polar ring behaves
as if they are continuous structures.
NGC 3034 (M82) shows a very exceptional behavior of rotation:
it has a steep nuclear rise as usual, but decreases after the
nuclear peak, obeying the Keplerian law.
This may have occurred by a strong tidal truncation of the disk,
by a close encounter with M81 (Sofue 1998).
NGC 4631 is an interacting dwarf galaxy with peculiar morphology, 
showing a rigid-body increase of rotation velocity.
Since this galaxy is edge on, it is not clear, either if the CO
gas is lacking indeed in the center, or it results from a true
rigid-body rotation.
NGC 4945 shows peculiar dark lanes and patches, but shows a quite 
normal rotation: steep nuclear rise and flat disk rotation.
Its peculiar distribution of interstellar gas is probably not due to 
its global dynamical characteristics, but due to a local phenomenon in 
the gas disk.

The interacting galaxy NGC 5194 (M51) shows a very peculiar rotation
curve (Fig. 4b, Fig. A1), which declines more rapidly than the 
Keplerian law at $R\sim 8 - 12$ kpc.
This may be due to variation of the inclination angle with the
radius, or warping.
In fact, this galaxy is nearly face-on ($i = 20^\circ$), but a slight warp
would cause a large error in deriving the rotation velocity:
If the galaxy's outer disk at 12 kpc has an inclination as small as
$i \sim 10^\circ$, such an apparently steep decrease would be observed
even for a flat rotation.

\centerline{ --- Fig.4  ---}

\subsection{Activity and Rotation Curves}

Our sample includes galaxies having various activities, such as
starbursts (NGC 253, NGC 1808, NGC 3034),
Seyferts (NGC 1068, NGC 1097),
LINERs (NGC 3521, NGC 4569, NGC 7331), and nuclear jets (NGC 3079).
In Figure 4c we plot rotation curves for these galaxies.
The global rotation and mass distribution appear rather normal
in these active galaxies, and no peculiar behavior is found
within our resolution.
This implies that such activities are triggered by a more local
and secondary cause than by a global dynamical mass distribution.
An exception is the starburst galaxy NGC 3034 (M82),
which shows a usual nuclear rise but is followed by a Keplerian
declining rotation, indicating a tidal truncation of the disk
(Sofue 1998), as mentioned in the previous subsection.

\section{UNIVERSAL PROPERTIES}

\subsection{Averaged Rotation Curves}

In Figure 5 we show mean rotation curves, which have been
obtained by averaging rotation velocities for three mass classes:
massive galaxies with a maximum disk velocity greater than 250 \kms;
galaxies with maximum disk velocities between 200 and 250 \kms;
and less-massive galaxies with velocities less than 200 \kms.
In Figure 5, we also plot the universal rotation curves (URC)
formulated by Persic et al (1996).
The observed mean rotation curves at $R>10$ kpc are well fitted by the URC.
Since the URC has been derived specifically for studying massive halos,
the disk and central rotation curves may not be well reproduced.
In fact, the central rotation curve obtained in this study is 
much steeper than the URC.
This may be partly due to the difference in the method to derive the 
rotation curve:
Rotation curves in the previous studies have been usually derived from the
intensity peak in the PV diagrams.
On the other hand, in the present study we used the
envelope-tracing method, which is thought to be more reliable to trace 
the central rotation curve (Sofue 1996; Takamiya and Sofue 1999b).
A relation discussion of rotation curves and position-velocity diagrams
will be given in subsection 4.2.

\centerline {--- Fig. 5  ---}

\subsection{Classification of Central Rotation Curves}

The observed rotation curves can be classified into the following
three types, according to their behavior in the central regions.
Figure 6a to 6c show the classified rotation curves.

{\it Central Peak Type}:
Rotation velocity attains a sharp maximum near the center at
$R\sim 100 - 500$ pc, followed by a dip at $\sim 1$ kpc, then
by a broad maximum of the disk component.
Examples are the Milky Way, NGC 891, NGC 3079, NGC 5907, and NGC 6946.

{\it No Central-Peak Type}:
The rotation curve rises steeply at the center, followed immediately by
a flat  part.
Examples are NGC 224, NGC 253, IC 342, NGC 5194, and NGC 5055.

{\it Rigid-Body Type}:
The rotation velocity increases mildly from the center in a rigid-body
fashion within the central 1 kpc.
This type is rather the exception, and is found in less-massive Sc-type 
galaxies, such as NGC 598, NGC 2403, NGC 3198, and NGC 4631.
This tendency has been already noticed by Casertano and van Gorkom (1991).

\centerline{--- Fig. 6 ---}

\subsection{Logarithmic Rotation Curves}

Since the dynamical structure of a galaxy varies with the radius
rapidly toward the center, an alternative plot, such as
in logarithm, of rotation curves would help to overview the
innermost kinematics.
In fact, the logarithmic plot in Figure 1b has demonstrated its
convenience for discussing the central mass distribution, including
the black hole of our Galaxy.
In Figure 7, we plot the same rotation curves as in Figure 2a in
logarithmic scaling.
The central 1 kpc regions are better presented in this figure.
However, resolutions of our data are not sufficient to express
rotation characteristics in the very center, particular within 100 pc for
many galaxies.

Although we must be careful in looking at this figure in the
sense that the resolution is limited in the central few hundred pc,
we may safely argue that high-mass galaxies show almost constant
rotation velocities from the center to outer edge,
in so far as they are presented in such a logarithmic plot.
On the other hand, lower-mass galaxies show decreasing rotation
toward the center  in the central 1 kpc regions.
However, we must also notice that the declining rotation toward
the center might be caused by insufficient angular resolution,
with which position-velocity diagrams often miss the central
steep rise.

\centerline{--- Fig. 7 ---}

A logarithmic rotation curve is particularly useful for such cases
with central massive black holes.
Figure 7b shows  rotation curves of four galaxies from our sample,
for which the existence of massive black holes is evident
(Richstone et al 1998):
NGC 4258 (Miyoshi et al 1995), Milky Way (Genzel et al 1997; Ghez et al 1998),
NGC 224 (Magorrian et al 1998) and
NGC 4945 (Greenhill et al 1997).
Here, equivalent rotation velocities corresponding to the
black hole masses are plotted by tilted straight lines, obeying
$V \propto R^{-1/2}$.
Except for the Milky way, they are connected to known rotation curves
by horizontal straight lines in the regions where detailed observations
are lacking.
In these galaxies, the rotation velocity never declines to zero
at the center.

\subsection{Scale Radius and Disk Rotation Velocity}

Maximum rotation velocity is an indicator of total
mass of a galaxy (e.g. Persic et al 1996).
The disk mass is also related to the optical scale radius ($h$), because 
the mass-to-luminosity ratio would not vary significantly inside the disk.
It is interesting to see how the scale radius of a disk is
correlated with the maximum velocity of rotation, which
will occur at $R=2.2 h$ for an exponential disk as is observed
in Fig. 2.
In Figure 8 we plot rotation velocities at $R=h$ and $R=2.2 h$
against the scale radius, $h$, from our sample galaxies,
where open circles are Sb, filled circles Sc,
and those with triangles are barred galaxies.

Although the correlation between velocity and scale radius
is not tight, there appear to exist two types of rough correlation's:
One group contains galaxies with rotation velocities
linearly correlated with the scale radius, approximately
fitted by the dashed line.
Another group comprises galaxies with largely-scattered
velocities but with scale radii being about constant at
$h \sim 2.3$ kpc.
No particular trend, such as larger scatter, for barred galaxies
is found in this plot.

\centerline{-- Fig. 8 --}

\section{DISCUSSION}

\subsection{Steep Nuclear Rise and High Rotation
Velocity in the Center}

The extremely high frequency of massive galaxies showing
the steep nuclear rise indicates that the high velocity is not due to a
particular view of non-circular motion by chance.
Note that the probability of looking at a bar end-on
is much smaller than that of viewing one side-on, which should result in
a larger probability for apparently slower rotation
than circular velocity at a given radius, and therefore, masses estimated
from the circular assumption would be even underestimated,
if they contain a bar.
In so far as our sample galaxies are concerned,
the steep nuclear rise of rotation velocity
is a universal property for massive Sb and Sc galaxies,
regardless of the existence of a bar and morphological peculiarities.
However, less-massive galaxies tend to show a rigid-body rise.

We may summarize that the rotation curves of massive Sb and 
Sc galaxies, in general, comprise the following four components:

{\parindent=0pt \parskip=0pt

(1) Steep central rise and peak, often starting from an already
high velocity at the nucleus;

(2) Bulge component;

(3) Broad maximum by the disk; and

(4) Halo component.
}

We stress that the rotation velocities in many well-resolved
galaxies do not decline to zero at the nucleus.
This indicates that the mass density increases toward the
nucleus more rapidly than expected from exponential or de Vaucouleur laws.
We mention that the widely adopted zero-velocity in the center
may be merely due to a custom of drawing a rotation curve by linking
positive and negative velocities from the opposite sides of the nucleus.

\subsection{High Velocities in the Center and Observed PV Diagrams}

We point out that nearer galaxies with higher effective resolution
in our sample tend to show steeper nuclear rise.
This suggests that an even steeper central rise would be observed in
more distant galaxies, if they were observed at higher resolution.
In order to examine how central high velocities would appear
in an observed PV diagram with finite resolution,
we have simulated PV diagrams from
assumed rotation curves and a model distribution of interstellar gas.
Figure 9a shows a case in which the rotation velocity is assumed to be 
constant, so that the velocity does not decline to zero at the center,
and the interstellar gas is centrally peaked.
However, the calculated PV diagram shows a rigid-body behavior near the
center, and if we trace the intensity peak on this PV diagram,
the velocity at the center would be measured to be zero.
Hence, in so far as a rotation curve constructed from a PV diagram is concerned,
we should not take its central rigid body-like behavior strictly:
the true velocity may be much higher, or may even start from a finite value
near a central black hole.

\centerline{ --- Fig. 9 ---}

Figure 9b shows a case with a more realistic rotation curve,
comprising four components: a central compact core, bulge, disk
and massive halo, each expressed by a Plummer potential.
In the calculated PV diagram, however, the central steep rise and
the peaks due to the core and bulge are hardly recognized.
This simulation demonstrates that tracing peak-intensity loci
in observed PV diagrams could miss possible central high velocities.
We, thus, conclude that the central rotation curves derived from observed
PV diagrams generally give {\it lower limits} to the rotation velocities.
In fact, high-resolution central PV diagrams observed for
some spiral galaxies by Bertola et al (1998) have indicated velocities
much higher than those currently supposed,
indicative of massive central black holes.
Detailed discussion of PV diagrams and a method to derive more
reliable rotation curves will be given in a separate paper 
(Takamiya and Sofue 1999b).

\noindent{\bf References}
\def\r{\hangindent=1pc  \noindent}

\r[Clemens 1985]{cle85} Clemens, D. P. 1985, ApJ 295, 422

\r Bertola et al 1998 

\r Bosma  A. 1981,  AJ  86,  1825 

\r Casertano  S.,  van Gorkom  J. H. 1991,  AJ 101,  1231

\r Genzel, R., Eckart, A., Ott, T., and Eisenhauer, F.
	1997, MNRAS 291, 219.

\r Ghez, A., Morris, M., Klein, B. L., Becklin,E.E. 
	1998, ApJ 509, 678.

\r Greenhill, L.J., Moran, J.M., Herrnstein, J.R.. ApJ 481, L23.

\r Honma, M., and Sofue, Y. 1997 PASJ 49, 453.

\r Honma, M., and Sofue, Y. 1997 PASJ 49, 539. 

\r Honma, M., Sofue, Y.,  Arimoto, N.  1995, AA 304, 1-10.

\r Kent,  S. M. 1987, AJ 93,  816.  

\r Kent,  S. M. 1991, ApJ 378, 131.	

\r Magorrian, J., Tremaine, S., Richstone, D., Bender, R., Bower, G.,
	Dressler, A., Faber, S.M., et al 1998 AJ, 115, 2285.

\r Mathewson, D.S. and Ford, V.L., 1996 ApJS, 107, 97.

\r Miyoshi, M., Moran J. Heernstein, J., Greenhill, L., Nakai, N.,
	Diamond, P., Inoue, M. 1995, Nature 373, 127.

\r Persic, M., Salucci, P., Stel, F.  1996, MNRAS, 281,  27.

\r Persic, M.,  and Salucci, P.  1995, ApJS 99, 501.

\r Richstone, D., Bender, R., Bower, G., Dressler, A., Faber, S.
	et al. 1998 Nature A14.

\r Rubin  V. C., Ford  W. K., Thonnard  N. 1980, ApJ  238, 471

\r Rubin, V. C., Ford, W. K., Thonnard, N. 1982, ApJ, 261, 439

\r Rubin, V., Kenney, J.D.P., Young, J.S. 1997 AJ, 113, 1250.

\r Sofue, Y. 1996, ApJ, 458, 120 (Paper I) 

\r Sofue, Y. 1997, PASJ, 49, 17 (Paper II)

\r Sofue, Y. 1998, PASJ, 50, 227

\r Sofue, Y., Honma, M., Arimoto, N. 1995, 1995, AA 296, 33-44.

\r Sofue, Y., Tomita, A.,  Honma, M.,Tutui, Y. and Takeda, Y.
	1998, PASJ 50, 427. (Paper IV)

\r Sofue, Y.,  Tutui, Y., Honma, M., and Tomita, A.,
	1997, AJ, 114, 2428 (Paper III)

\r Takamiya, T., and Sofue, Y. 1999a, b, in preparation.

\r van der Kruit, P.C., and Searle, L. 1982 AA 110 61.  

\r Warner, P. J.; Wright, M. C. H., and  Baldwin, J. E
       1973, MNRAS, 163, 163.

\newpage

\noindent{\bf Appendix}

We present rotation curves for individual galaxies in Figure A1.
Machine-readable data in the form of tables are available
by contacting the first author.
Table A1 lists the parameters of the galaxies.

\centerline{--- Figure A1 ---}

\centerline{--- Table 1 ---}

\newpage

\parindent=0pt
Figure Caption

Fig. 1. (a) Rotation curve of the Milky Way plotted in
linear scale. (b) Rotation curve plotted
in a logarithmic scale, which obeys the Keplerian law
near the center due to the massive black hole.

Fig. 2. Most-completely-sampled rotation curves  of
Sb and Sc galaxies obtained by using CO, \ha\ and HI-line data.
(a) All galaxies;
(b) All galaxies for central 5 kpc;
(c) Rotation curves with radii normalized by
scale radius of exponential disk.

Fig. 3. (a) Sb;  (b) Sc; and
(c) barred galaxies (SBb and SBc).

Fig. 4. (a) Peculiar and Irregular galaxies, 
(b) interacting galaxies, and 
(c) galaxies with active nuclei and/or starbursts.

Fig. 5. Mean rotation curves for galaxies with disk rotation
velocities higher than 250 \kms, between 200 and 250 \kms, and
lower than 200 \kms.
Universal rotation curves (URC) formulated by Persic et al (1996)
are shown by dashed lines for several values of galaxy luminosities.

Fig. 6. Classification of rotation curves into three types
according to the central behavior.

Fig. 7. Logarithmic rotation curves:
(a) All galaxies; (b) Galaxies with central massive black holes.

Fig. 8. Rotation velocities at scale radius $h$ and $2.2h$ plotted
against $h$. Open circles are Sb, filled circles Sc, and those with
triangles are barred galaxies.

Fig. 9. Simulation of position-velocity diagrams from assumed
rotation curves and gas distribution:
(a) Rotation velocity is assumed to be constant, and therefore,
the central velocity is finite.
(b) Four mass components (core, bulge, disk and halo) are assumed,
having a very sharp central rise of rotation velocity.
Note that the high velocities in the center and steep nuclear rise
are hardly detected in PV diagrams with finite resolution.
This simple simulation suggests that observed PV diagrams often
miss a massive compact component in the center.

Fig. A1. Individual rotation curves for spiral galaxies.

\newpage

\def\tableline{\hline}
\begin{table*}
\caption{List of galaxies}
\begin{tabular}{ccccccccccc}
\tableline \tableline
 Name     &RA(1950)&Dec(1950)&Type&Act&Int.&PA(deg)&Incl.(deg)& $D$ (Mpc) &$h$ (kpc) \\
\tableline
 Milky Way&          &           & Sb  &   &   &    &   90    &  0        &         &   \\
 NGC 224  & 00 40 00 & +40 59 43 & Sb  &   &   & 40 &   77    &  0.69     &  5.3    &   \\
 NGC 253  & 00 45 06 & -25 33 40 & Sc  &SB &   & 51 &   78.5  &  2.5      &  2.3    &   \\
 NGC 598  & 01 31 02 & +30 24 15 & Sc  &   &   & 22 &   54    &  0.79     &  2.7    &   \\
 NGC 660  & 01 40 21 & +13 23 25 & Sc  &   &   & 45 &   70    & 13        &         &   \\
 NGC 891  & 02 19 25 & +42 07 19 & Sb  &   &   & 19 &   88.3  &  8.9      &  4.2    &   \\
 NGC 1003 & 02 36 06 & +40 39 28 & Scd &   &   & 97 &   66    &  9.52&         &   \\
 NGC 1068 & 02 40 07 & -00 13 32 & Sb  &Sy &   & 79 &   46    & 18.1      &         &   \\
 NGC 1097 & 02 44 11 & -30 29 01 & SBb &Sy &   &135 &   40    & 16        &         &   \\
 NGC 1365 & 03 31 42 & -36 18 27 & SBb &   &   &222 &   46    & 15.6      &  5.8    &   \\
 NGC 1417 & 03 39 28 & -04 51 50 & Sb  &   &   &175 &   50    & 54.1      &         &   \\
 IC 342   & 03 41 59 & +67 56 26 & Sc  &   &   & 40 &   25    &  3.9      &         &   \\
 UGC 2855 & 03 43 16 & +69 58 46 & SBc &   &   &112 &   61    & 28.77     &  4.6    &   \\
 NGC 1808 & 05 05 59 & -37 34 37 & Sbc &SB &   &138 &   58    & 11.4      &         &   \\
 UGC 03691& 07 05 11 & +15 15 33 & Scd &   &   & 65 &   65    & 30.0      &         &   \\
 NGC 2403 & 07 32 06 & +65 42 40 & Sc  &   &   &125 &   60    &  3.3      &  2.13   &   \\ 
 NGC 2590 & 08 22 29 & -00 25 42 & Sb  &   &   & 77 &   71    & 64.5      &  2.1    &   \\
 NGC 2708 & 08 53 37 & -03 10 05 & Sb  &   &   & 20 &   68    & 24.6      &         &   \\
 NGC 2841 & 09 18 36 & +51 11 24 & Sb  &   &   &150 &   68    &  9        &  2.3    &   \\ 
 NGC 2903 & 09 29 20 & +21 43 19 & Sc  &   &   & 21 &   35    &  6.1      &  1.9    &   \\
 NGC 3031 & 09 51 27 & +69 18 08 & Sb  &   &   &152 &   59    &  3.25     &  2.5    &   \\
 NGC 3034 & 09 51 44 & +69 55 01 & I   &SB & y & 71 &$\sim 90$&  3.25     &         &   \\
 NGC 3079 & 09 58 35 & +55 55 15 & Sc  &jet&   &169 &$\sim 90$& 15.6      &         &   \\
 NGC 3198 & 10 16 52 & +45 48 06 & SBc &   &   &215 &   70    &  9.1 &  2.5    &   \\
 NGC 3495 & 10 58 41 & +03 53 43 & Sd  &   &   & 20 &   85    & 12.8      &         &   \\
 NGC 3521 & 11 03 16 & +00 14 11 & Sbc &LIN&   &166 &   75    &  8.9  &  2.4    &   \\
\tableline
\label{table1}
\end{tabular}
Note: The data are taken from Papers I-IV. The scale radii have been
adopted from Kent (1987), van der Kruit and Searle (1982), and Honma and
Sofue (1997).
\end{table*}

\begin{table*}
\begin{tabular}{ccccccccccc}
\tableline \tableline
 Name     &RA(1950)&Dec(1950)&Type&Act&Int.&PA(deg)&Incl.(deg)& $D$ (Mpc) &$h$ (kpc) \\
\tableline
 NGC 3628 & 11 17 49 & +13 51 46 &Sb/I &   &   &104 & $>86$   &  6.7      &         &   \\
 NGC 3672 & 11 22 30 & -09 31 12 & Sc  &   &   &  8 &   67    & 28.4      &         &   \\
 NGC 3953 & 11 51 12 & +52 36 18 & SBc &   &   & 13 &   54    & 20.70     &         &   \\
 NGC 4062 & 12 01 31 & +32 10 26 & Sc  &   &   &100 &   68    &  9.7      &         &   \\
 NGC 4096 & 12 03 29 & +47 45 20 & Sc  &   &   & 20 &   73    & 12.22     &         &   \\
 NGC 4258 & 12 16 29 & +47 34 53 & Sbc &   &   &150 &   67    &  6.6      &  5.6    &   \\
 NGC 4303 & 12 19 22 & +04 45 03 & Sc  &   &   &318 &   27    &  8.1 &         &   \\
 NGC 4321 & 12 20 23 & +16 05 58 & Sc  &   &   &146 &   27    & 15   &         &   \\
 NGC 4448 & 12 25 46 & +28 53 50 & SBab&   &   & 94 &   71    &  9.7      &         &   \\
 NGC 4527 & 12 31 35 & +02 55 44 & Sb  &   &   & 66 &   69    & 22        &         &   \\
 NGC 4565 & 12 33 52 & +26 15 46 & Sb  &   &   &137 &   86    & 10.2      &  5.6    &   \\
 NGC 4569 & 12 34 19 & +13 26 16 & Sab &LIN&   & 23 &   63    &  8.2      &         &   \\ 
 NGC 4605 & 12 37 48 & +61 53 00 &SBc,p&   &   &125 &   69    &  4.0      &         &   \\
 NGC 4631 & 12 39 40 & +32 48 48 &Sc/I &   & y & 86 &   84    &  4.30     &         &   \\
 NGC 4736 & 12 48 32 & +41 23 32 & Sab &   &   &108 &   35    &  5.1      &         &   \\
 NGC 4945 & 13 02 32 & -49 12 02 &Sc/I &   &   & 43 &   78    &  6.7      &         &   \\
 NGC 5033 & 13 11 09 & +36 51 31 & Sc  &   &   &179 &   62    & 14   &  6.0    &   \\
 NGC 5055 & 13 13 35 & +42 17 39 & Sbc &   &   &103 &   55    &  8   &  3.8    &   \\
 NGC 5194 & 13 27 46 & +47 27 22 & Sc  &   & y & 22 &   20    &  9.6      &         &   \\
 NGC 5236 & 13 34 12 & -29 36 42 & SBc &   &   & 45 &   24    &  8.9      &         &   \\
 NGC 5457 & 14 01 26 & +54 35 18 & Sc  &   &   & 38 &   18    &  7.2      &         &   \\ 
 NGC 5907 & 15 14 36 & +56 30 45 & Sc  &   &   &156 &   88    & 11.6 &  6.0    &   \\
 NGC 6674 & 18 36 31 & +25 19 55 & SBb &   &   &143 &   55    & 42.62     &         &   \\
 NGC 6946 & 20 33 49 & +59 58 49 & Sc  &   &   & 64 &   30    &  5.5      &         &   \\
 NGC 6951 & 20 36 37 & +65 55 46 & Sbc &   &   &154 &   48    & 18        &         &   \\
 NGC 7331 & 22 34 47 & +34 09 21 & Sbc &LIN&   &167 &   75    & 14        &  4.7    &   \\
\tableline \tableline
\label{table1}
\end{tabular}
{Note: SB=starburst; LIN=LINER; Sy=Seyfert; Int=Interaction; y=yes.  }
\end{table*}

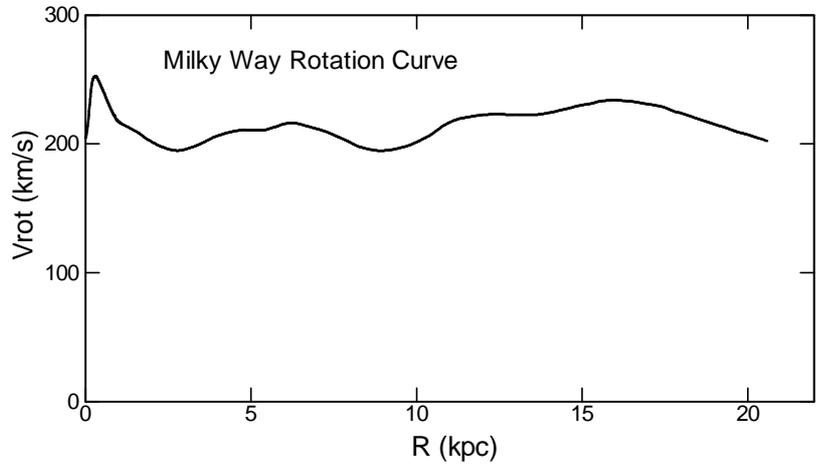
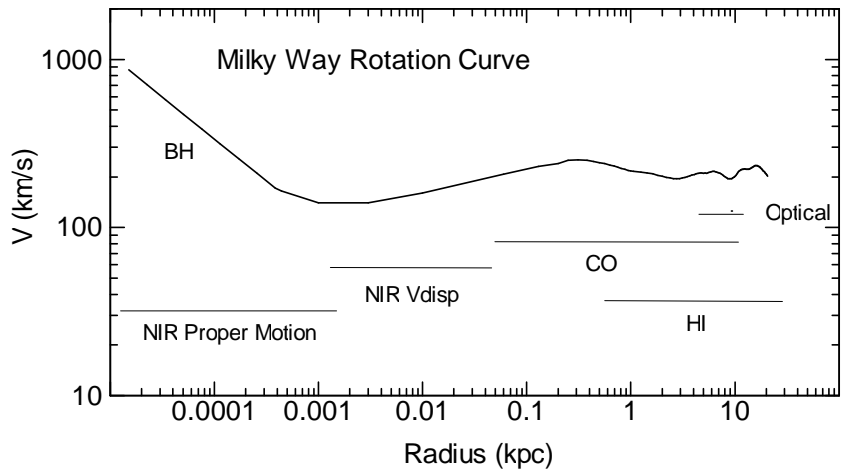
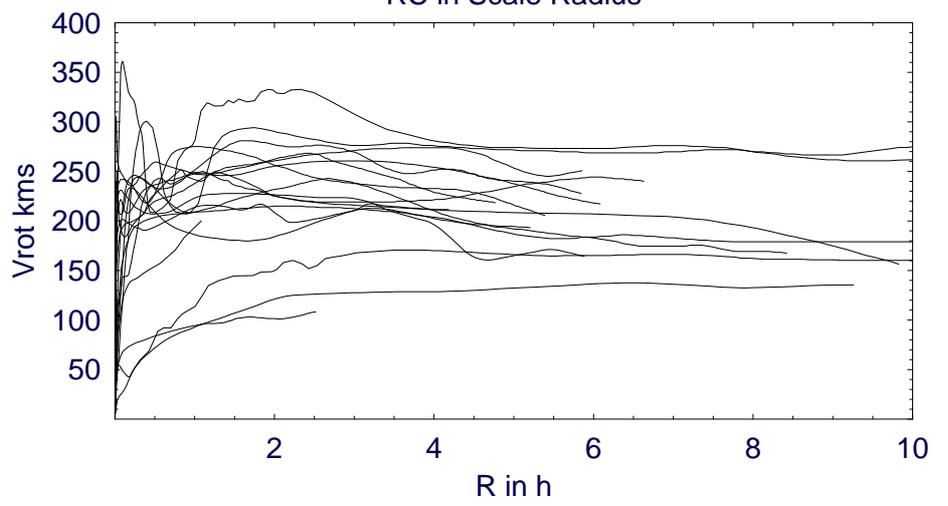
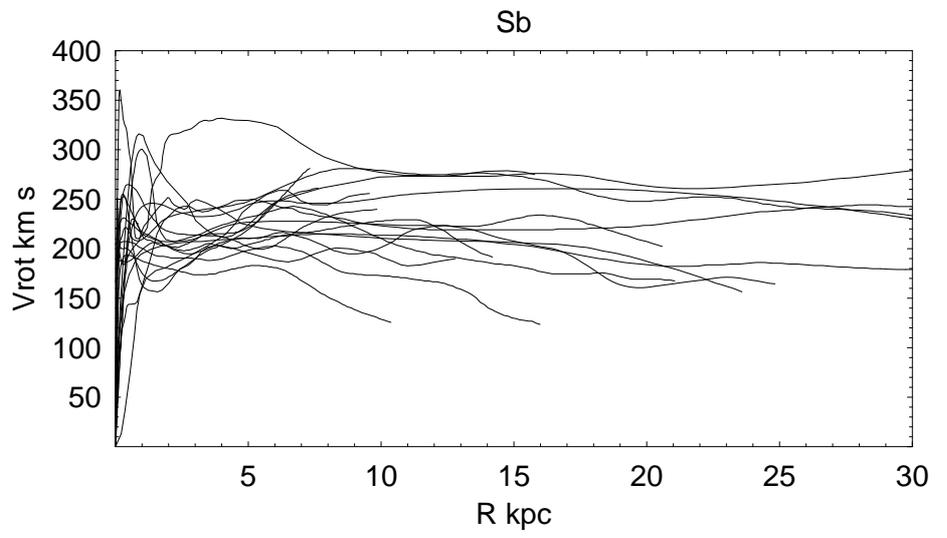
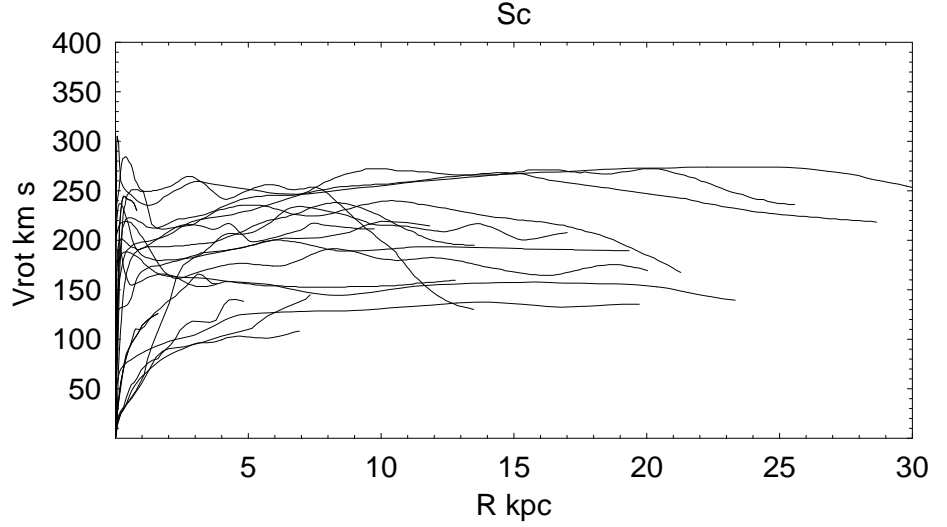
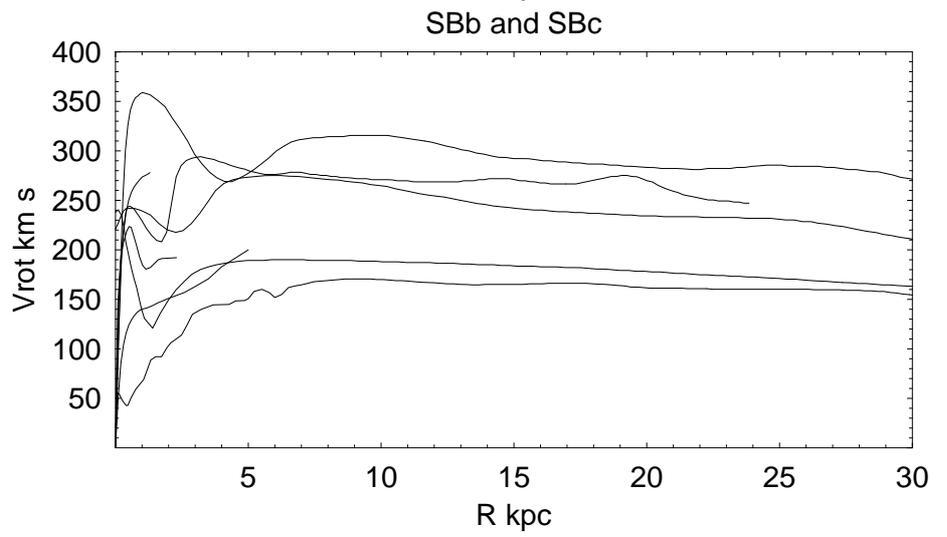
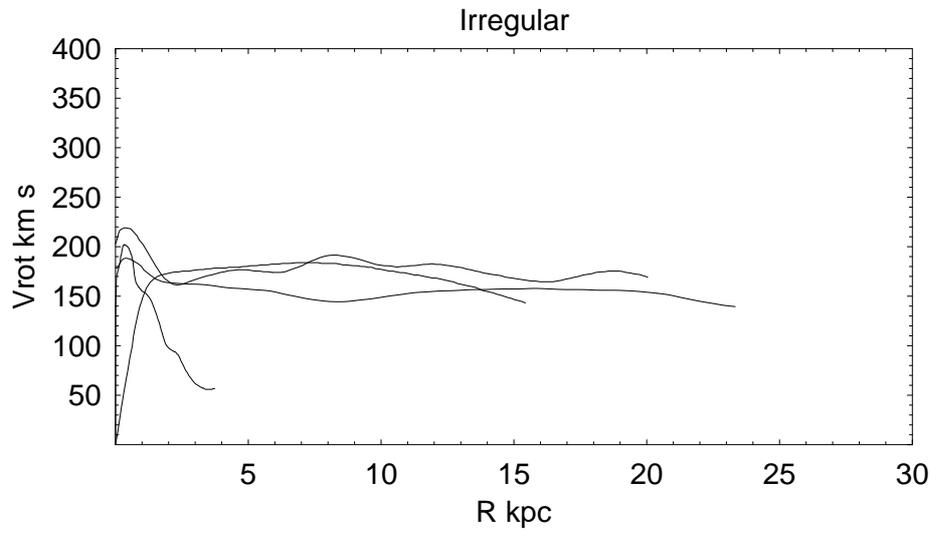
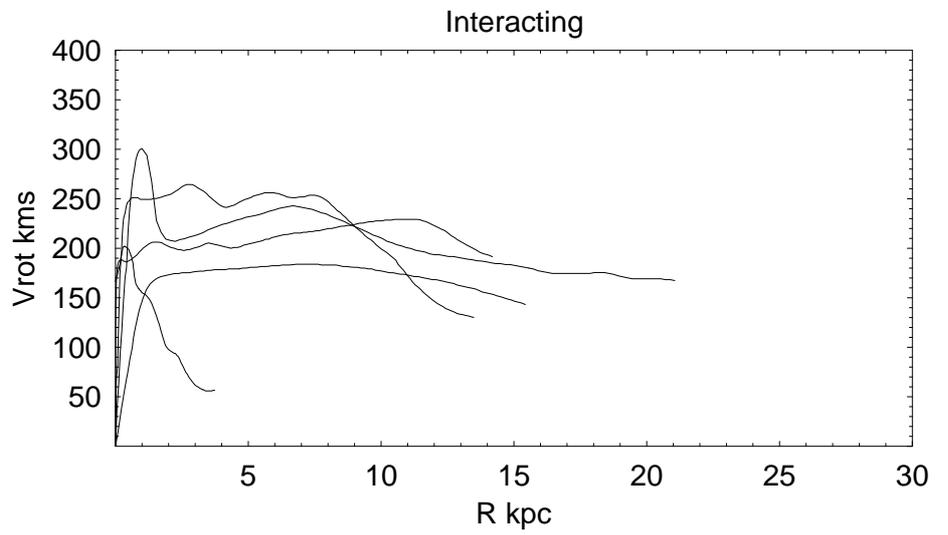
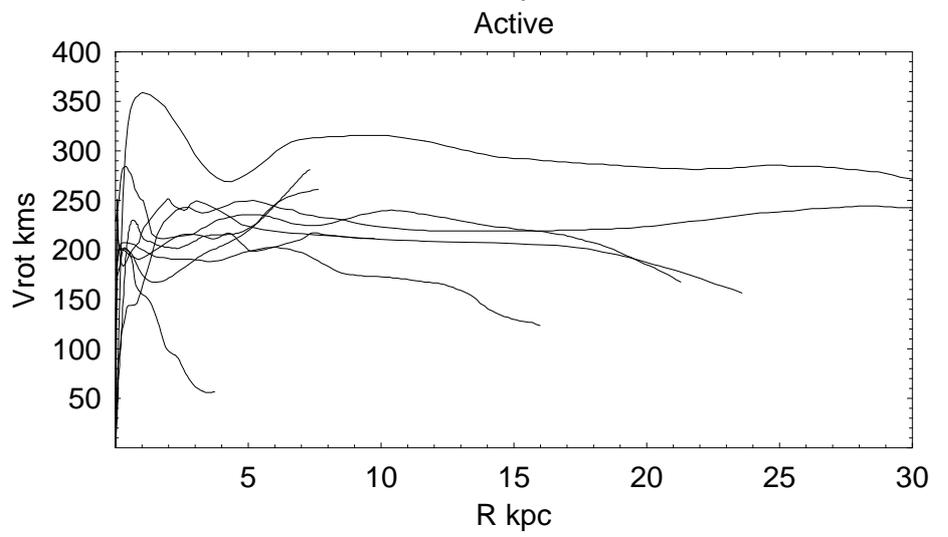
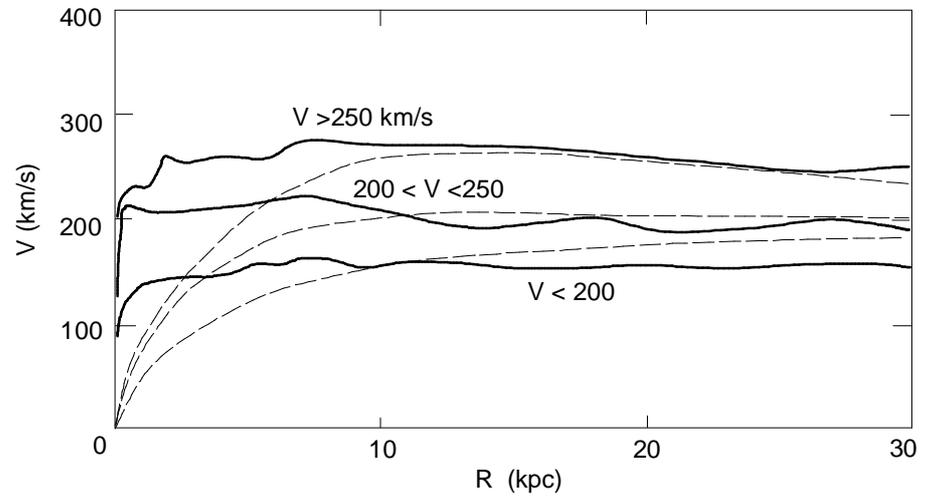
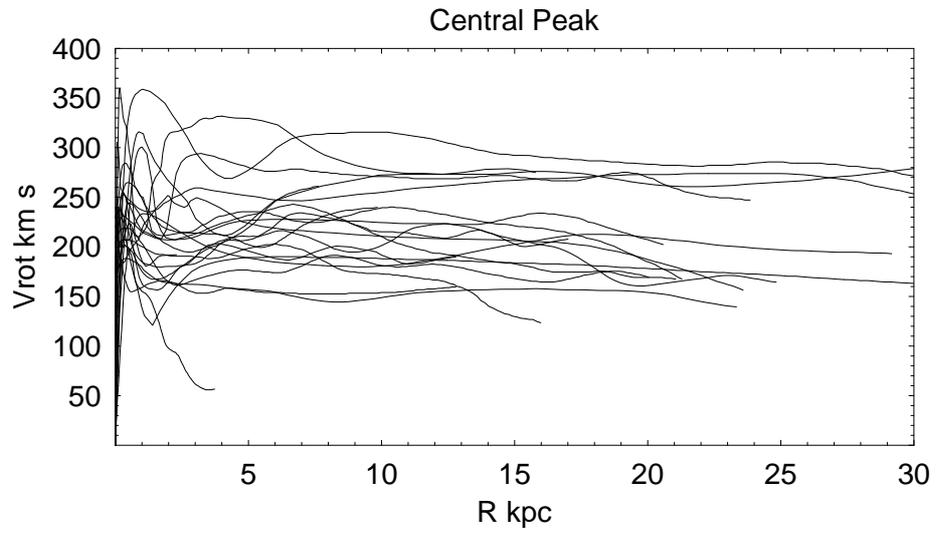
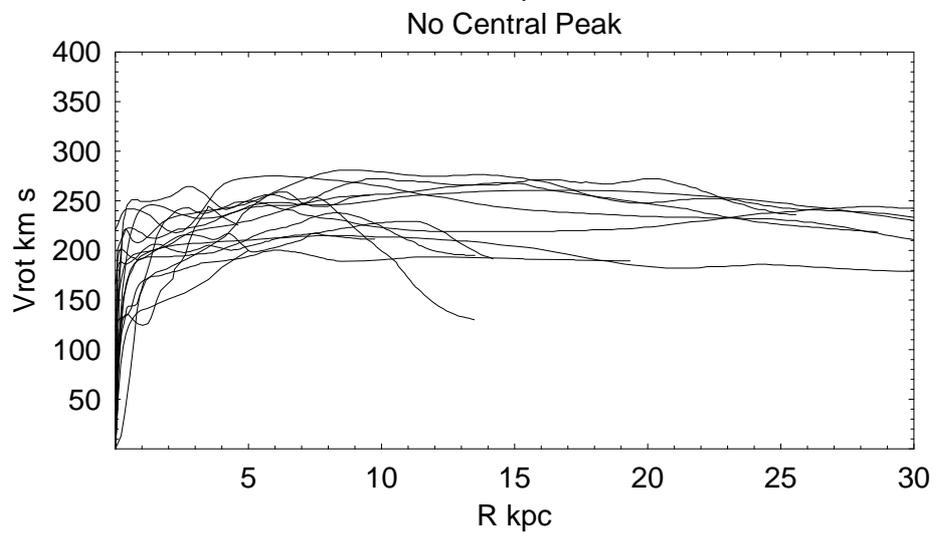
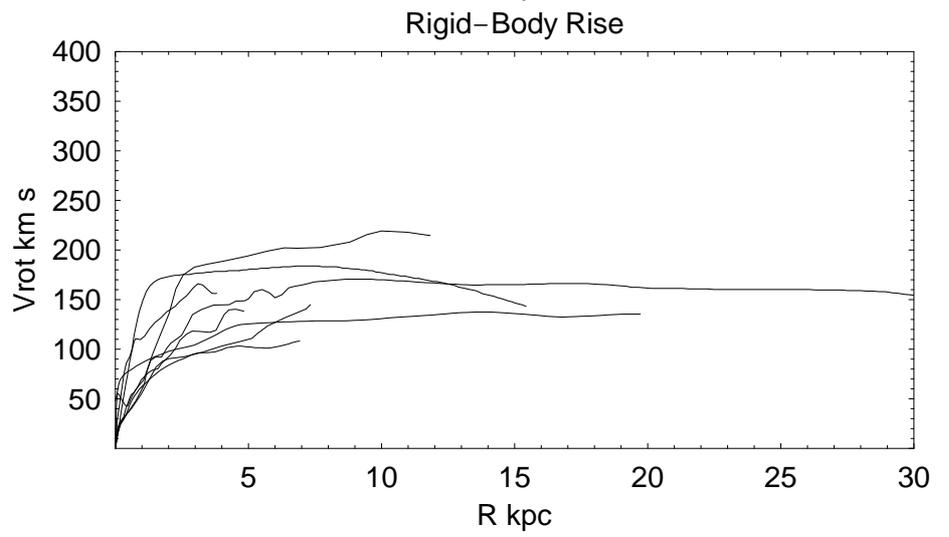
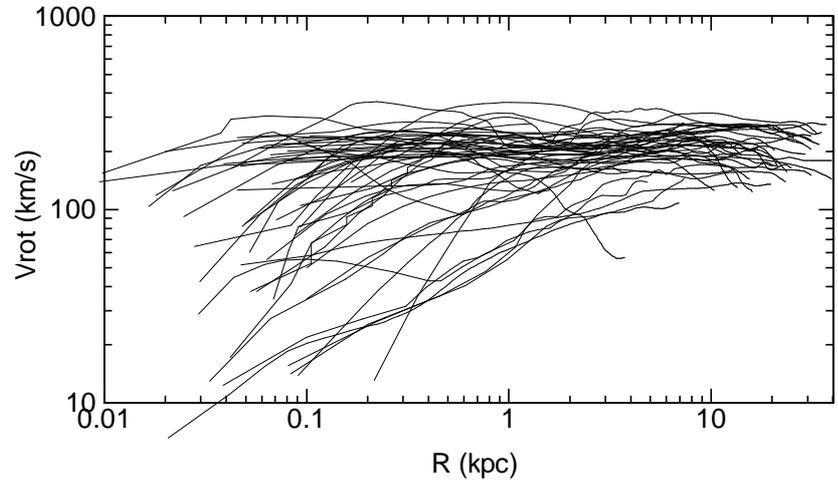
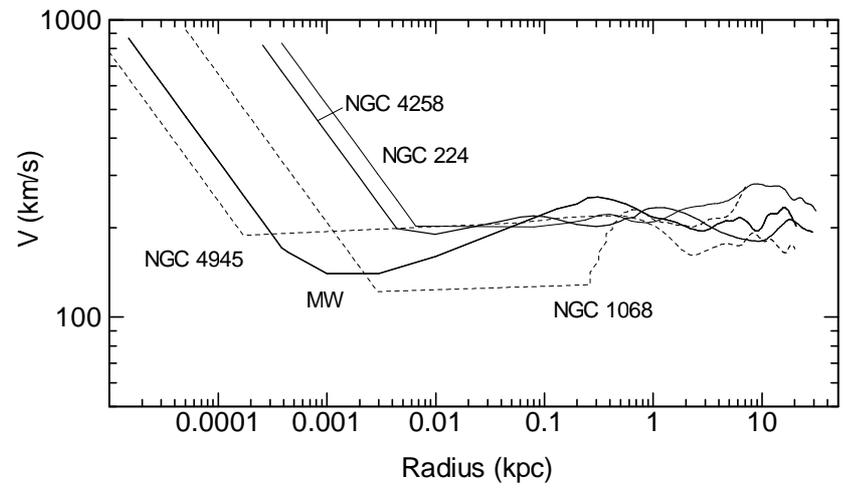
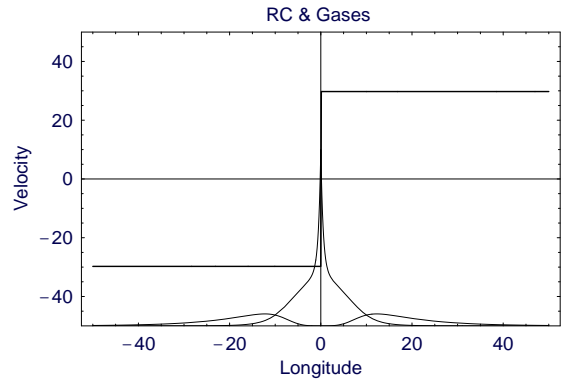
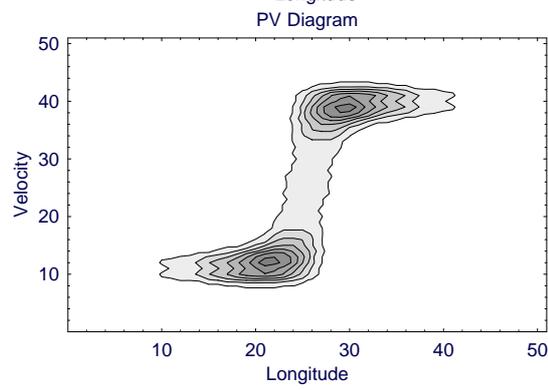
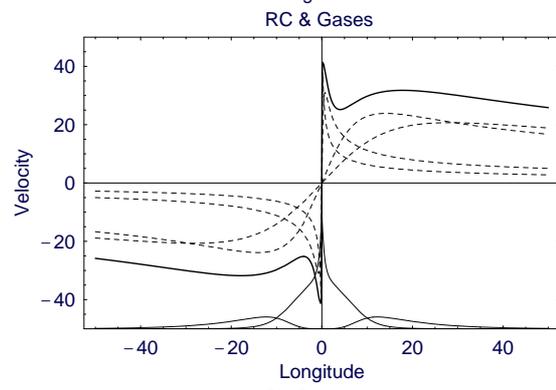
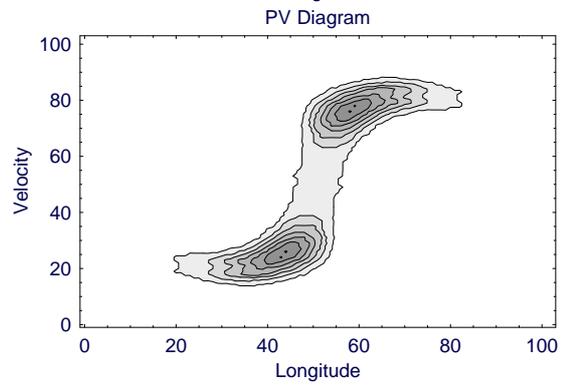
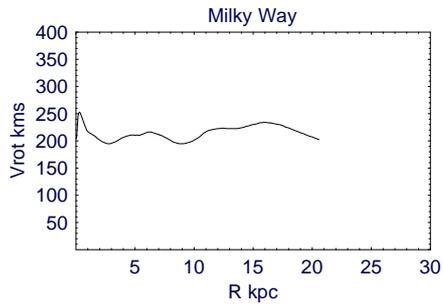
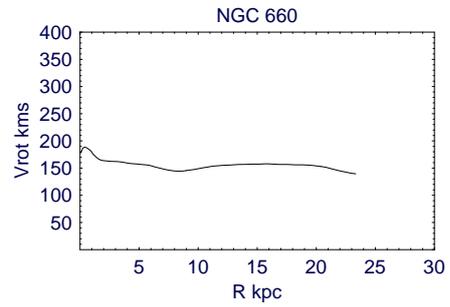
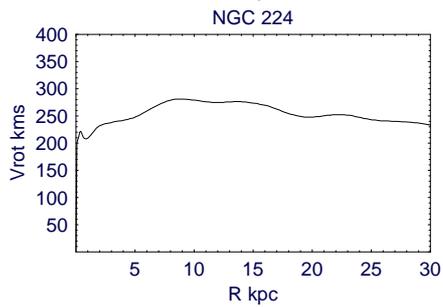
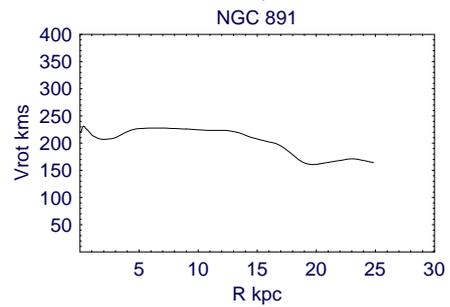
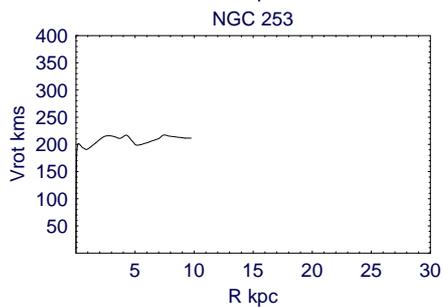
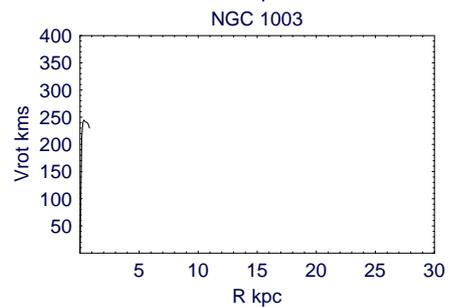
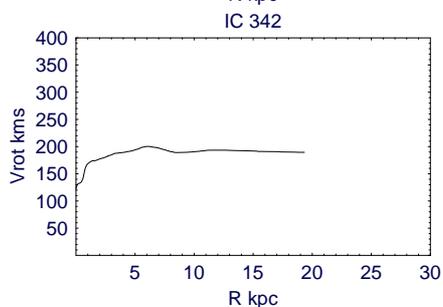
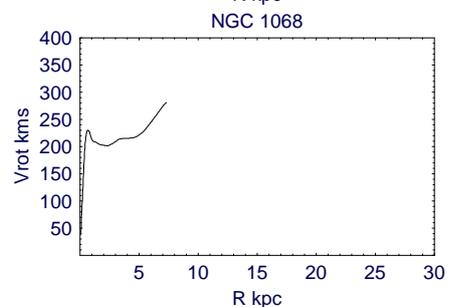
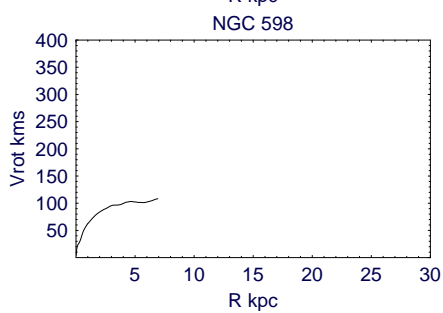
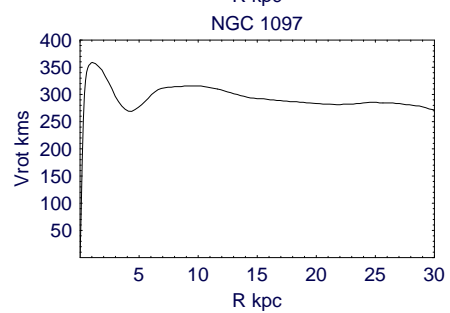
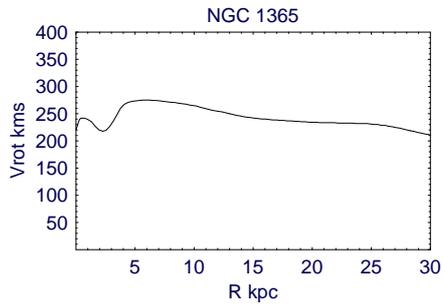
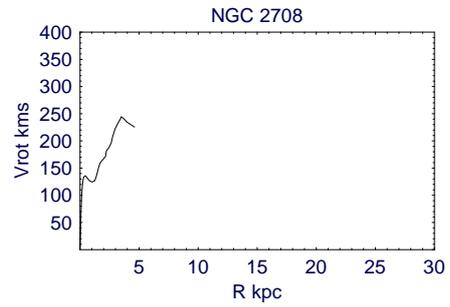
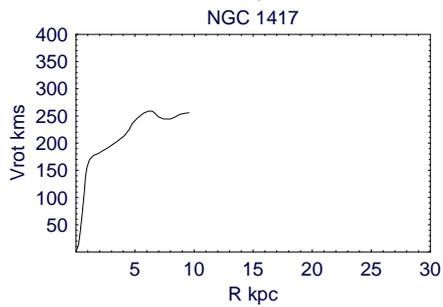
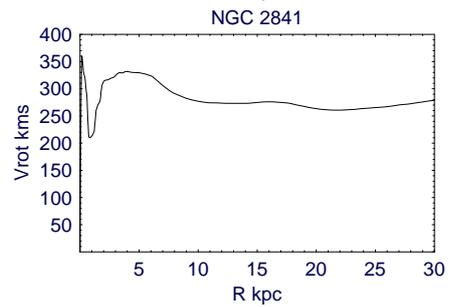
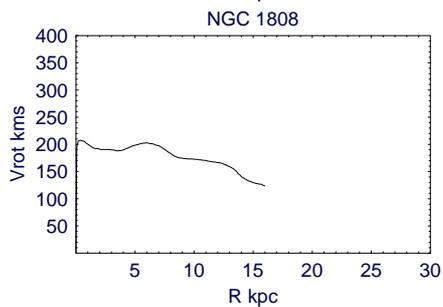
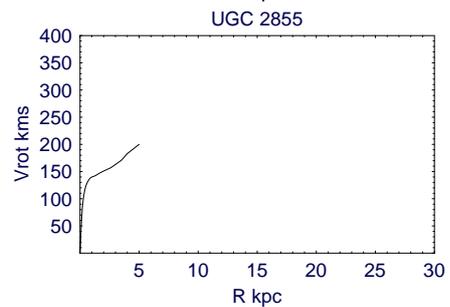
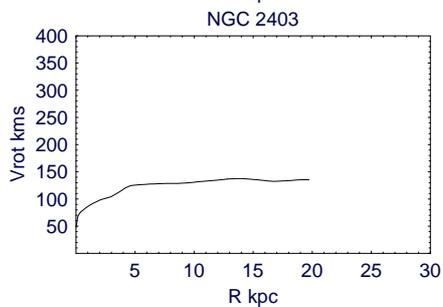
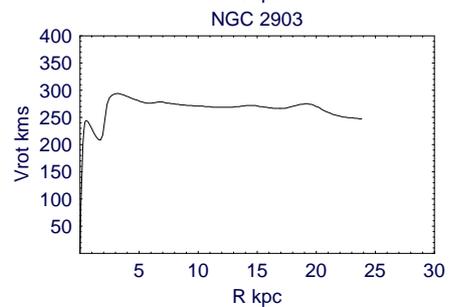
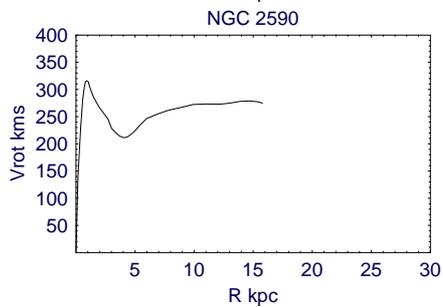
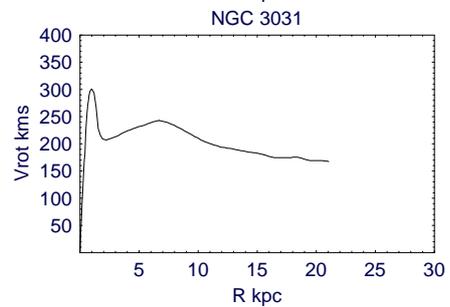
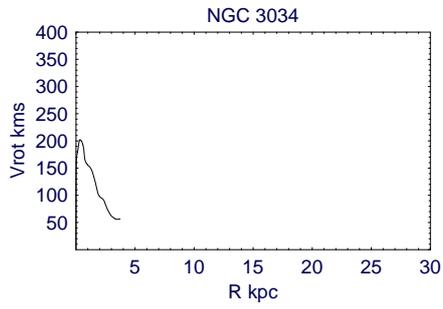
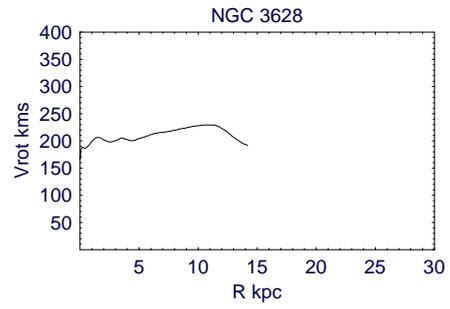
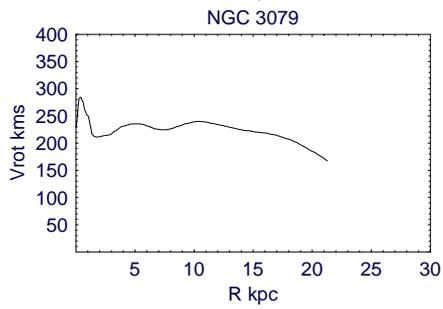
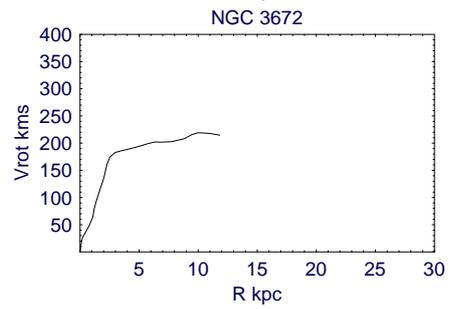
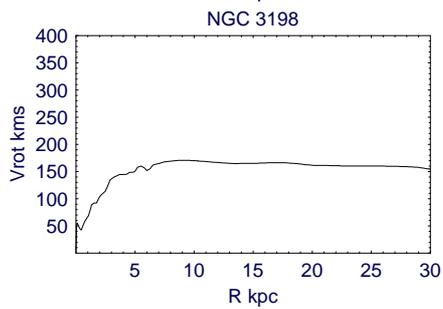
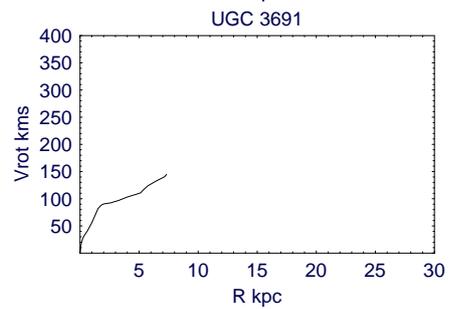
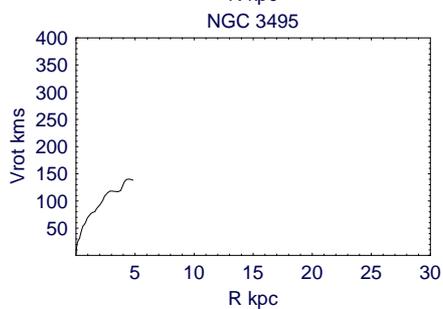
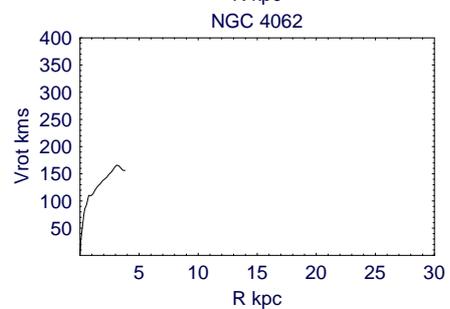
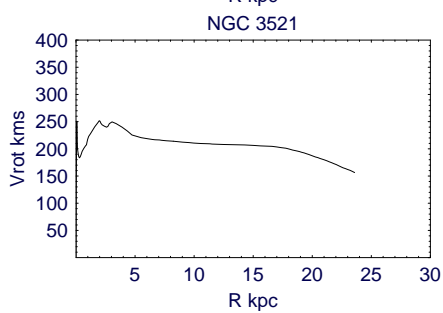
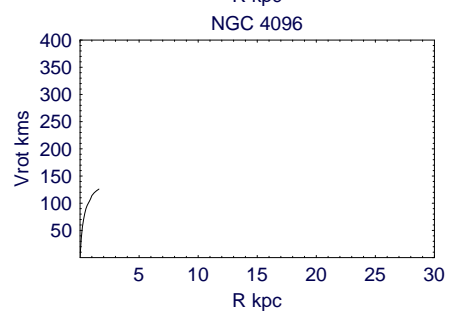
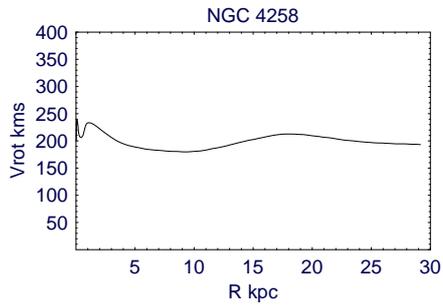
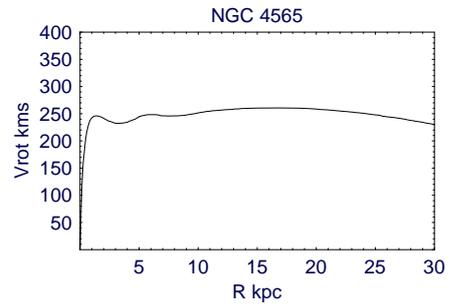
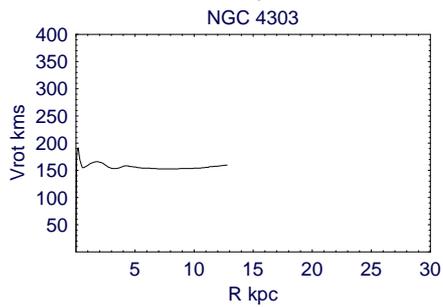
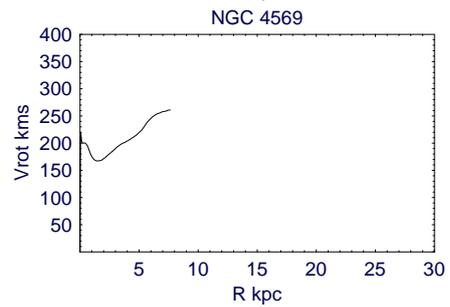
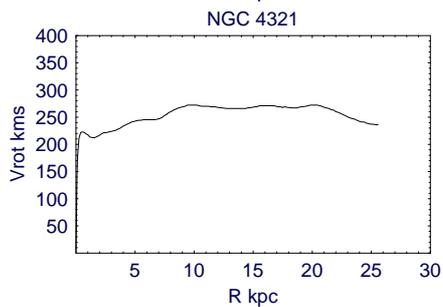
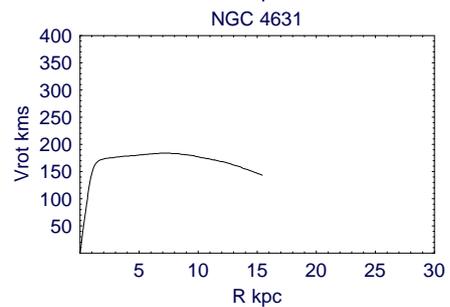
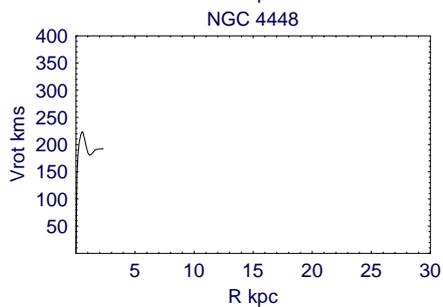
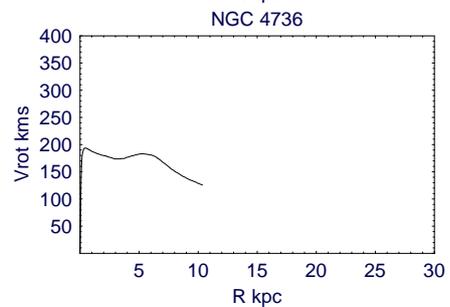
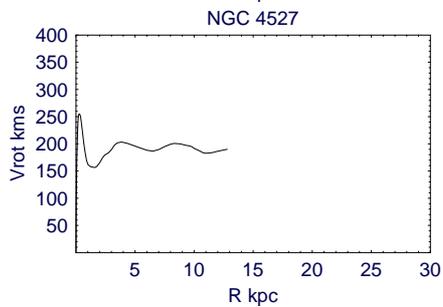
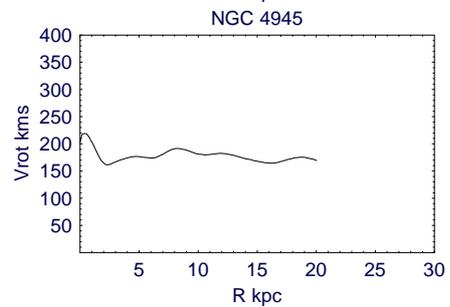
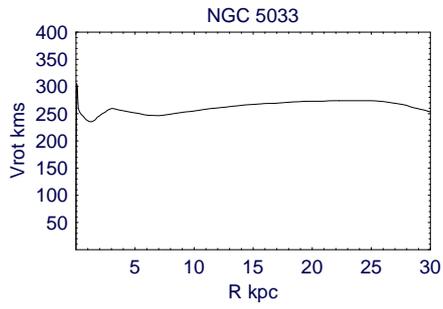
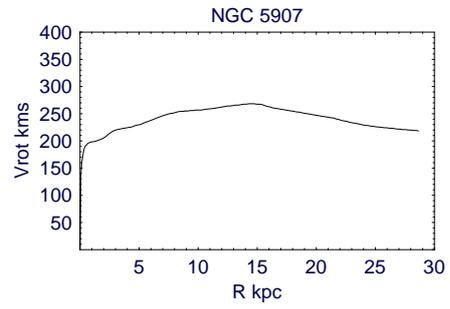
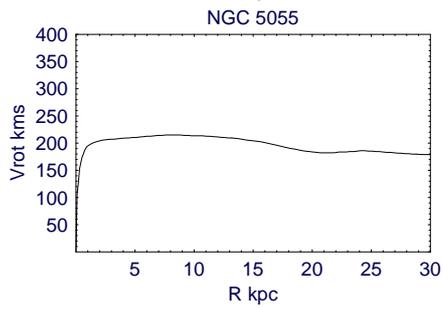
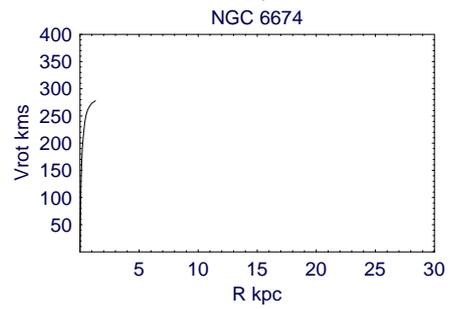
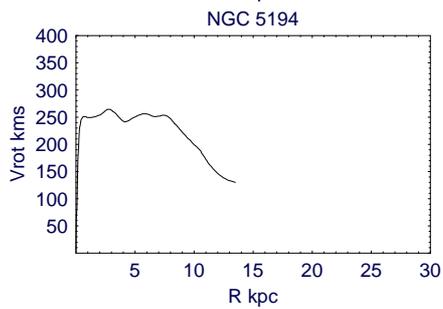
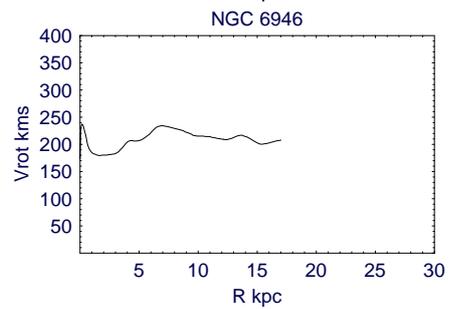
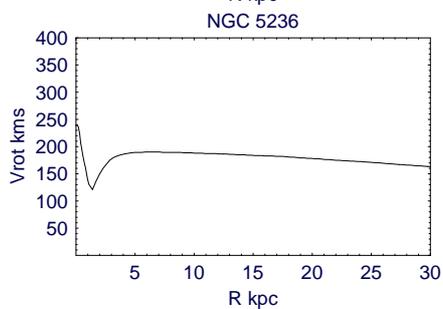
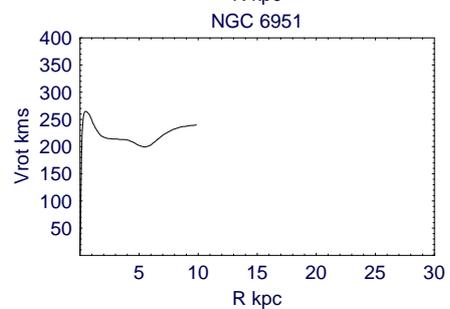
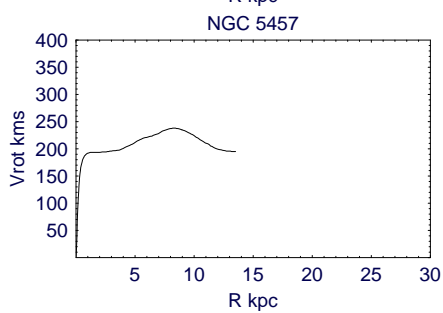
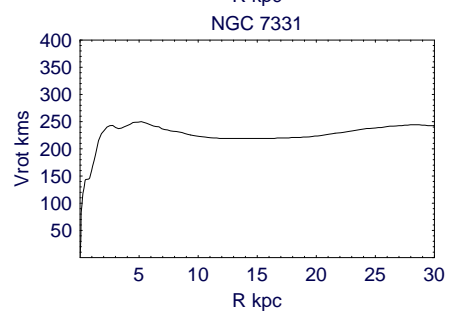
\end{document}